\documentclass[final,3p,twocolumn,times]{elsarticle} 
\usepackage{graphicx,latexsym}

\def\bea{\begin{eqnarray}}
\def\eea{\end{eqnarray}}
\def\be{\begin{equation}}
\def\ee{\end{equation}}
\def\ba{\begin{array}}
\def\ea{\end{array}}

\def\eps{\epsilon}
\def\MT{\mathcal{T}}
\def\BT{\mathbf{T}}
\def\BL{\mathbf{\Lambda}}
\def\Bx{\mathbf{\lambda}}

\def\bc{\begin{center}}
\def\ec{\end{center}}
\def\ds{\displaystyle}  

\def\mm#1#2#3#4{%
\left(
\begin{array}{lr}
\ds{#1}  &  \ds{#2}  \\
\ds{#3}  &  \ds{#4}
\end{array}
\right)
}

\def\mv#1#2{%
\left(
\begin{array}{l}
#1    \\
#2
\end{array}
\right)
}

\def\bS{\mathbf{S}}

\def\oA{\overline{A}}
\def\oB{\overline{B}}

\def\eps{\varepsilon}

\journal{}

\begin{document}

\begin{frontmatter}

\title{Propagation of the surface plasmon polaritons through gradient index and periodic structures}

\author{Tom\'a\v s V\'ary}
\author{Peter Marko\v s}
\address{Department of Physics FEI STU Bratislava, Slovakia
}

\begin{abstract} 
We study the propagation of surface electromagnetic waves 
along the metallic surface covered by various  layered dielectric structures.
We   show that  strong radiative losses typical for the scattering of 
the surface wave can be considerably suppressed when single dielectric step is substituted by gradient index or periodic
layered  structure.
\end{abstract}



\end{frontmatter}

\section{Introduction}

It is well known that the life time of surface electromagnetic waves  (surface plasmon polaritons - SPP) propagating along   the metal-dielectric interface
\cite{economou,zayats,meier} is rather short.
Even in the case of ideal lossless metal, the propagation of the  SPP through any inhomogeneity is accompanied with radiative  losses  
\cite{krenn}.  The simplest example  is a transmission of the SPP
through a  single dielectric step, represented by the planar interface between two dielectrics
which cover the metallic surface \cite{lahm,stegeman}.
As was shown recently \cite{oulton,vm}, SPP can radiate 
up to 40 per cent of its energy at the single interface. 

The origin of radiative  losses lies in the different spatial distribution of the electric and magnetic 
field in the direction
perpendicular to the metallic surface. To balance the tangential
components of $\vec{E}$ and $\vec{H}$, the SPP radiates plane waves. 

The radiative losses can be suppressed for SPP propagating at the interface between metamaterials.
The losses are almost negligible for the TE polarized 
SPP propagating along the surface of negative permeability material. The physical reason for this is
that the spatial distribution of the TE polarized SPP depends only weakly on the permittivity $\epsilon_d$
of the dielectric \cite{VaryEtopim}. In  Ref. \cite{prl}, anisotropic metamaterials were proposed to reduce 
radiative losses for the TM polarized SPP. 

The presence  of plane waves makes the quantitative analysis of the propagation of SPP mathematically difficult. 
Analytical calculation of the transmission and reflection coefficients \cite{stegeman} of the SPP through  
discontinuities represents 
rather difficult task.  Numerical methods \cite{oulton,vm}  are very useful for the
quantitative analysis of the scattering. The problem is formulated for the 
 scattering matrix of both SPP and radiative plane waves.
With $N_w$ plane waves considered, the size of the scattering matrix increases to $2(1+N_w)\times 2(1+N_w)$.

In this paper, we analyze the propagation of the SPP along the metallic surface covered by various 
dielectric materials. Of special interest is the propagation through layered structure when the metal is covered by parallel
dielectric strips of various with and permittivities. We consider  gradient index structures, in which the permittivities in two adjacent 
strips differs in small value $\Delta\eps$, and periodic structures composed of alternating  dielectrics.
We calculate the transmission and reflection coefficients for the SPP and show that  radiative losses  can be 
considerably reduced when the single interface between two dielectrics is substituted by the layered structure
in which the permittivity smoothly changes. 

The paper is organized as follows.
In Section \ref{def} we define the model and introduce the parameters of the SPP. Section \ref{t}
discuss the  the method of calculation of the transmission parameters for SPP propagating through
the layered structure.
In Section \ref{grad} we show that  radiative losses can be dramatically 
reduced when  the single interface $\epsilon_1/\epsilon_2$ is 
substituted by the gradient index structure \cite{kuzmiak}, 
represented by $N_w$ planar layers with linearly increasing 
dielectric permittivity.
In Section \ref{slab} we discuss the  propagation of the SPP through planar slab.
Similarly to the case of the single interface, we find much higher transmission when the slab is modeled by the
gradient structure (sect. \ref{q}).
Propagation of the SPP through  layered medium constructed by periodic alternation of two dielectrics
$\epsilon_1$ and $\epsilon_2$ is analyzed in Section \ref{pc}. 
We show that the propagation of SPP through such structure does not 
require an increase of radiative losses: radiation is the same, or even smaller, 
than that accompanying the propagation through single interface.

\section{Definition of the model}\label{def}

Consider the metallic surface lying  in the  $xy$ plane with metal filling the $z<0$ region. 
The metal is covered by different dielectrics. All interfaces
between two neighboring dielectrics are  parallel to  the $yz$ plane. The SPP propagates along the
metallic surface. The direction of the propagation is perpendicular to dielectric interfaces.
The electric and magnetic field  of the SPP 
decreases exponentially in the $z$ direction as $\exp(-\kappa_{d}z)$ for $z>0$ (dielectric) and $\exp(+\kappa_mz)$
for $z<0$ (metal).
The analytical form of parameters $\kappa_d$ and $\kappa_m$ is \cite{economou}
\bea\label{spp-k}
\kappa_d^2 = - k_0^2\ds{\frac{\eps_d^2}{\eps_d+\eps_m}}~~~~\textrm{and}~~~~ 
\kappa_m^2 = - k_0^2\ds{\frac{\eps_m^2}{\eps_d+\eps_m}}. 
\eea
Here, 
$k_0=\omega/c$ and $c$ is the light velocity,
$\eps_m$ is a permittivity of metal, represented by lossless Drude formula,
$\eps_m  = 1 - \omega_P^2/\omega^2$ and  $\eps_d$ is dielectric constant of 
dielectric medium. 
The  SPP propagates along the $x$ direction.
Then, its wave vector in the $xy$ plane has only   $x$-component,
\be\label{spp-kx}
k_{x}^2 = k_0^2\ds{\frac{\eps_d\eps_m}{\eps_d+\eps_m}}.
\ee
Equations (\ref{spp-k},\ref{spp-kx}) give complete frequency dependence of the wave vector of the  SPP.

For the metal-dielectric interface only the TM polarized SPP can be excited  \cite{wp}.
The intensity of the magnetic  field is
\be \label{spp-h}
\vec{h} = \left\{
\begin{array}{ll}
{\cal{N}}_0(0,1,0)~\Phi(x,y,t)~ e^{-\kappa_dz} & z>0, \\
{\cal{N}}_0(0,1,0)~\Phi(x,y,t)~ e^{+\kappa_mz} &z<0, 
\end{array}\right.
\ee
and the intensity of the electric field is 
\be \label{spp-e}
\!\!\!\!\!\!\!\vec{e}  = \left\{
\begin{array}{ll}
{\cal N}_0\ds{\frac{z_0}{k_0}}~ \Phi(x,y,t)~ (+i\kappa_d, 0, -k_{x})e^{-\kappa_d z}/\eps_d  
& z>0,\nonumber \\
{\cal N}_0\ds{\frac{z_0}{k_0}}~ \Phi(x,y,t)~ (-i\kappa_m, 0, -k_{x})e^{+\kappa_m z}/\eps_m &
 z<0.
\end{array}\right.
\ee
Here,   $\Phi(x,y,t) = e^{i(k_xx-\omega t)}$ and $z_0=\sqrt{\mu_0/\epsilon_0}$. The  constant
${\cal N}_0$  normalizes the energy current in the $x$ direction \cite{sch,oulton}.

Similar formulas can be derived for the  plane waves scattered at the metallic interface. Explicit expression for
the electric and magnetic fields  is given elsewhere \cite{oulton,vm}. Here we only note that  we will
consider $N_w$ plane waves with the same
frequency $\omega$ but different  $z$-component of the wave vector $k_z$:
\be
k_{z\alpha}=\ds{\frac{k_{z {\rm max}}}{N_w}}\alpha,~~~~~~\alpha=1,2,\dots N_w.
\ee
Both $N_w$ and $k_{z {\rm max}}$  represent the parameters of the model.
Other components of the plane wave vectors are $k_y=0$ and
\be\label{kxa}
k_{x\alpha}=\ds{\sqrt{k_0^2\epsilon_d-k_{z\alpha}^2}}.
\ee
Only waves  with real $k_{x\alpha}$ contribute to radiative losses.

In what follows we consider the frequency of the SPP
\be\omega=0.23\omega_p,
\ee
which corresponds to the frequency of visible light, and neglect absorption.
Then, the  metallic permittivity is  $\eps_m = -17.9$. 
Using Eq. \ref{spp-k}, we express  the wavelength $\Lambda$ of the SPP:
\be\label{Lambda}
\Lambda_{\epsilon_d} =\ds{ \lambda_p\frac{\omega_p}{\omega}\sqrt{\frac{\eps_d+\eps_m}{\eps_d\eps_m}}}.
\ee
where $\lambda_p=2\pi c/\omega_p\approx  94.2$~nm is the wavelength corresponding to the plasma frequency
$\omega_p/2\pi = 2\times 10^{15}$ s$^{-1}$.

\section{Propagation of the  SPP through planar interfaces }\label{t}

Scattering of SPP on the single interface is accompanied by the radiation losses \cite{vm}. Therefore, the 
scattering matrix $T$ is of the size of $2(N_w+1)\times 2(N_w+1)$, where N is a number of plane waves, given by the 
$z$-component of the wave vector. The amplitudes of  scattered waves are  related  by the scattering matrix 
\be
\mv{B}{\oA}=\mm{\bS_{11}}{\bS_{12}}{\bS_{21}}{\bS_{22}}\mv{\oB}{A}.
\ee
Here, $A_i$ and $B_i$ are the amplitudes of waves on the left and right side of the interface.
Amplitudes $\oA_i$ and $\oB_i$ belong to the plane waves propagating in the opposite direction.
We have $N_w+1$ waves,   ($i = 0 \dots N_w$)  with
$A_0$  and
$B_0$ representing  the amplitudes of the SPP.

The scattering matrix expresses the waves propagating away from the interface ($B$, $\oA$) through the amplitudes of the
 incident waves ($A$, $\oB$).
Explicit form of scattering matrix is given in \cite{oulton,vm}. 
The transmission and the reflection coefficients of the SPP incident from medium $1$ are given by
\be
T_{1\to 2} = |\bS_{12}(00)|^2, ~~~~~~
R_{1\to 1} = |\bS_{22}(00)|^2,
\ee
and from medium 2
\be
T_{2\to 1} = |\bS_{21}(00)|^2, ~~~~~~
R_{2\to 2} = |\bS_{11}(00)|^2.
\ee
Radiative losses into the first and second media are
\bea
S_{1\to 1} = \sum_{\alpha}^{N_1} |\bS_{22}(\alpha 0)|^2,  ~~~
S_{1\to 2} = \sum_{\beta}^{N_2} |\bS_{12}(\beta 0)|^2, \\
S_{2\to 2} = \sum_{\alpha}^{N_1} |\bS_{11}(\alpha 0)|^2,  ~~~
S_{2\to 1} = \sum_{\beta}^{N_2} |\bS_{21}(\beta 0)|^2,
\eea
respectively. Here, $N_i$  is a number of propagating plane waves (waves with real $x$ component of the wave vector)
in the $i$th medium.
Conservation of energy requires
\be
T_{1\to 2} + R_{1\to 1} + S_{1\to1} + S_{1\to2} = 1,
\ee
since the absorption is omitted.
In this way we calculate transmission parameters of the surface wave propagating through the
single planar interface. For more complicated structures which 
contain two or more interfaces, we apply the transfer matrix method.
To every interface we assign it's transfer matrix defined as

\be
\BT_{BA}
=\left( \ba{ll}
        S_{12} - S_{11} S_{21}^{-1} S_{22} & S_{11} S_{21}^{-1}
        \\ -S_{21}^{-1} S_{22}  & S_{21}^{-1}
        \ea \right).
\ee 
The transfer matrix 
for the homogeneous dielectric spacing between two interfaces is
\be
\BL = \left( \ba{ll}
        \Bx & 0
        \\ 0 & \Bx^{-1}
        \ea \right).
\ee
Here $\Bx$ is a diagonal matrix containing phase factor $e^{i k_{x\alpha} L}$, where $L$ 
is the width of the dielectric slab. The
transfer matrix of the entire structure containing $n$ interfaces is given as a product
of successive transfer matrices  
\be\label{T}
\MT = \BT_{nn-1} \BL_{n-1} \BT_{n-1n-2} \dots \BL_2 \BT_{21} \BL_1 \BT_{10}.
\ee
In the process of multiplication we have to keep in mind that some of the elements of the 
matrices $\Bx_i^{-1}$ correspond to waves that increases exponentially 
along the $+ x$ direction.  
Therefore the elements of $\MT$ can become very large. To avoid numerical operations with
$\Bx^{-1}$ we apply the normalization procedure \cite{PMcKR}:
consider the  product  $\MT=\BT'\BL\BT$. To calculate the transmission coefficient,
we need 
the matrix $\MT_{22} = S_{21}^{-1}$ which can be expressed as
\be
\MT_{22} = \left( 0 \: 1 \right)
\left( \ba{ll}
        \MT_{11} & \MT_{12}
        \\ \MT_{21} & \MT_{22}
        \ea \right) 
\left( \ba{l}
        0 \\ 1
        \ea \right). \nonumber
\ee
This expression can be written in the form
\be
\MT_{22}=
(\BT'_{21}\: \BT'_{22})\mv{\Bx\BT_{12}\BT^{-1}_{22}\Bx}{1}\Bx^{-1}\BT_{22}
\ee
so that
\be
S_{21} = \MT_{22}^{-1}=T_{22}^{-1}\Bx(\BT_{21}'\Bx\BT_{12}\BT_{22}^{-1}\Bx + \BT_{22}')^{-1}, 
\ee
and the transmission coefficient of the SPP   $T = |S_{21}(0,0)|^2$ 
is  expressed without any use of elements of  $\Bx^{-1}$. 
Similarly, we express matrices  $\MT_{12}$ and  $S_{11} = \MT_{12}S_{21}$.
\be
\begin{array}{ll}
S_{11} = & \left( \BT_{11}' \Bx \BT_{12} \BT_{22}^{-1} \Bx + \BT_{12}' \right)\\
        &\times (\BT_{21}' \Bx \BT_{12} \BT_{22}^{-1} \Bx + \BT_{22}')^{-1}
\end{array}
\ee
which determines the reflection coefficient $R$.
This approach can be applied recursively to the multilayer structures. 
We calculate $S_{11}^{(n+1)}$ and
$S_{21}^{(n+1)}$ matrix elements for $(n+1)$-th interface by
\bea
S_{21}^{(n+1)} &= \!\!\!\!\!\!\!& S_{21}^{(n)}\lambda Y^{-1}   \nonumber\\
\!\!\!\!\!\!\!\!\!\!\!\!\!\!
S_{11}^{(n+1)} &= & \lambda X Y^{-1} \lambda,
\eea
where
\be
X = T_{11} S_{11}^{(n)} + T_{12},~~~~~~~ 
Y = T_{21} S_{11}^{(n)} + T_{22},
\ee
and $S_{11}^{(n)}$, $S_{21}^{(n)}$ are scattering matrices for structure up to $n$-th interface.


\begin{figure}[t!]
\begin{center}
\includegraphics[clip,width=11cm]{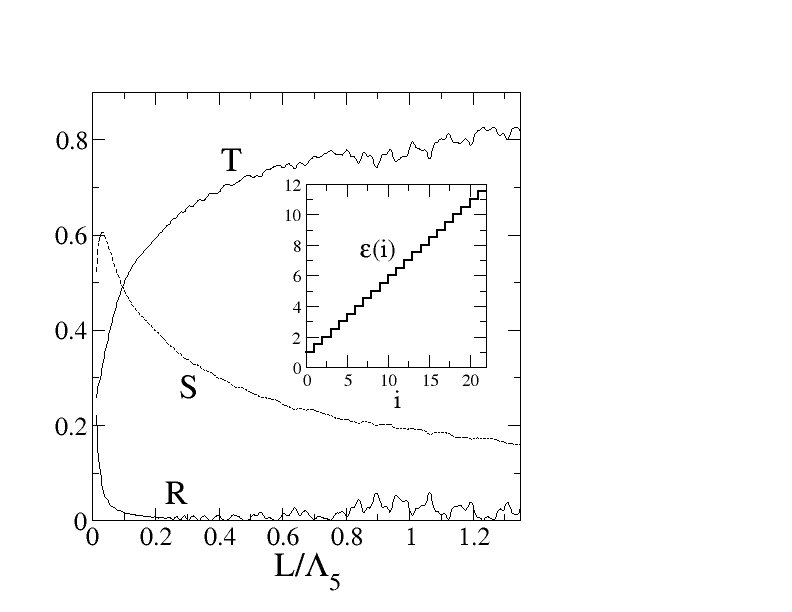} 
\end{center}
\caption{The transmission coefficient $T$, the reflection coefficient $R$ and radiative  losses $S$  
for the normal incident the SPP propagating through the gradient structure with   
linearly increasing value of the  permittivity. 
The permittivity changes from $\eps_A = 1$ to $\eps_B = 11.5$ with the increments $\Delta\eps = 0.5$
(spatial dependence of the permittivity is shown in the inset).
The width $L$ of each layer is measured relative to the wavelength $\Lambda_5$ of the SPP.  
The transmission and scattering losses for the single
interface  $\epsilon_A/\epsilon_B$ are $T= 0.19   $ and $S= 0.5 $, respectively.
}
\label{fig1}
\end{figure}
\begin{figure}[t!]
\begin{center}
\includegraphics[clip,width=11cm]{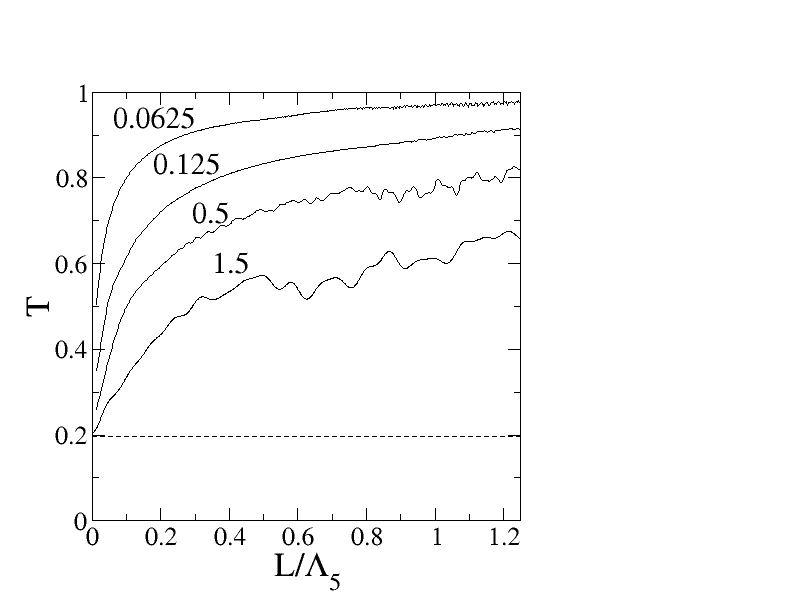}
\end{center}
\caption{The transmission coefficient $T$
for the normal incidence of the SPP propagating through the
layered structure
$\eps_i = \eps_A+\Delta\eps \times i$ for various permittivity increments $\Delta\eps$.
Horizontal dashed line represents the  transmission coefficient $T=0.19$  for the  single interface between media $\eps_A = 1$ 
and $\eps_B = 11.5$.}
\label{fig2}
\end{figure}

\section{Gradient structures}\label{grad}

Previous scattering analysis \cite{lahm,stegeman,oulton,vm} showed that the propagation of the SPP
 through the single dielectric interface with high index contrast is accompanied by 
huge scattering losses. To suppress these losses we analyze layered structure which consists of
$N$ parallel layers of the same width. The permittivity of the $i$-th layer is
\be\label{epsi}
\eps_i = \eps_A + \Delta\eps \times i, 
\ee
and the permittivity step  $\Delta \eps = (\eps_B - \eps_A)/N$.
This model mimics the gradient structure where the permittivity increases linearly from 
$\eps_A$ to $\eps_B$. 

Figure \ref{fig1} shows the transmission and the
 reflection coefficients  and an amount of scattering losses for 
the normal incidence SPP propagating  through such structure for 
$\eps_A = 1$ to $\eps_B = 11.5$. The structure is composed from 
$N=20$ dielectric layers of the same width. 
The transmission increases when the width of layers increases and saturates 
for sufficiently wide dielectric slabs. This effect is more significant for finer permittivity 
steps (Fig. \ref{fig2}).
It is obvious that radiative losses could be totally eliminated for perfectly smooth permittivity profile.
We also see that the reflectivity of this structure is very small  (Fig. \ref{fig1}). 

Of course, 
the limit of unity transmission coefficient  can be reached only theoretically.
With 20 permittivity steps,
considered in Fig. \ref{fig1}, 
the entire structure becomes macroscopically large.
when  the width of layers is $L=\Lambda_{5}\sim 245$~nm. 
However, as shown in Fig. \ref{fig2}, considerable increase of the 
transmission coefficient can be obtained already with much thinner layers and/or with the use of only 
a few dielectric layers.


\begin{figure}[t!]
\begin{center}
\includegraphics[clip,width=11cm]{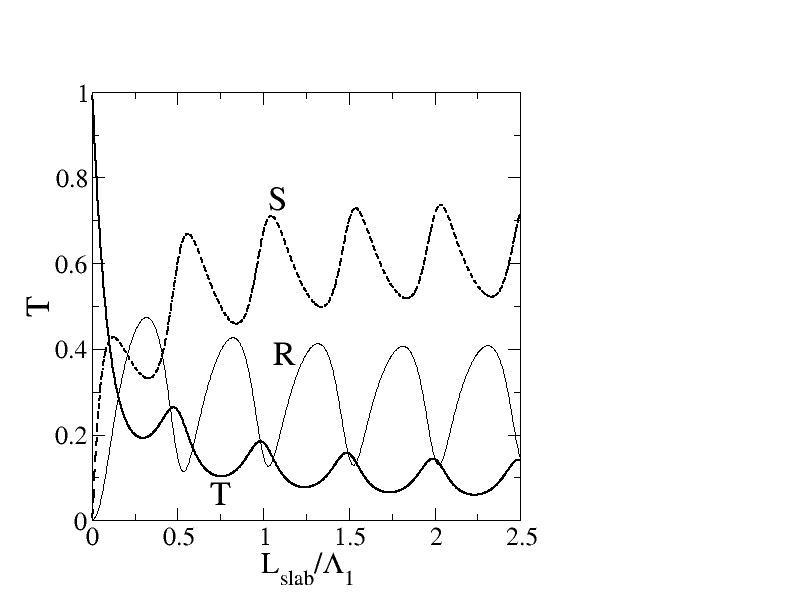} 
\end{center}
\caption{The transmission parameters  of the normal incidence of the 
SPP propagating through the dielectric slab with the permittivity $\eps_B = 1$ 
sandwiched between two dielectric media with the permittivity $\eps_A = 8$.  The thickness of the  slab
is shown relative to wavelength of the SPP $\Lambda_1$. 
Fabry-Perot resonances of both the transmission and the reflection are clearly visible. Note that
maxima of the transmission coefficient are accompanied with maxima of the radiative losses.
}
\label{fig51}
\end{figure}

\begin{figure}[t!]
\begin{center}
\includegraphics[clip,width=11cm]{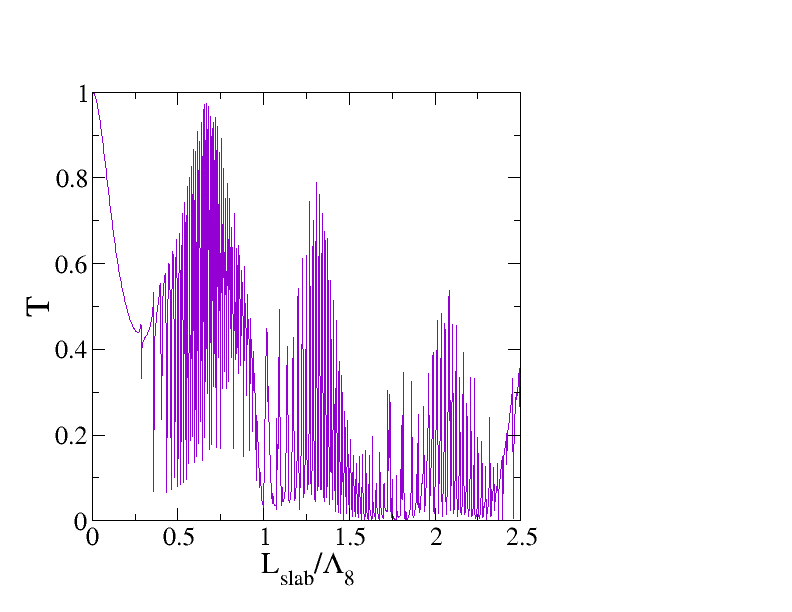} 
\end{center}
\caption{The transmission coefficient $T$, the reflection coefficient $R$ and scattering losses 
for the  propagation of the  SPP through the dielectric slab with the permittivity $\eps_B = 8$ 
sandwiched between dielectric media with the permittivities $\eps_A = 1$.  The thickness of the 
is shown relative to wavelength of the SPP $\Lambda_8$. 
}
\label{fig51a}
\end{figure}

\section{Propagation of the SPP through dielectric slab}\label{slab}

Consider now the  slab of dielectric material of width $L_{\rm slab}$ with permittivity $\eps_B$
embedded into two semi-infinite layers with permittivity $\eps_A$ 
and study the propagation of the SPP across such structure. 
The transfer matrix (\ref{T}) reduces to $\MT = \BT_{AB}\BL_B \BT_{BA}$.

Figure \ref{fig51} shows the transmission of SPP through the slab with $\eps_A=8$ and $\eps_B=1$.
The transmission and reflection coefficients exhibit typical
Fabry-Perot resonances which confirm that the transmission is determined mostly by the ratio
$L_{\rm slab}/\Lambda_1$  of the slab width  to the wave length of the SPP.
Similarly as in the case of the plane waves \cite{yeh,wp}
with maximal values for $L=n \times \Lambda_1/2$
and  maxima of the reflection $R$ appear when $L\approx (\Lambda_1/4)\times (2n+1)$. 
As shown in Fig. \ref{fig51}, the maxima of the transmission coefficients 
 are accompanied with maxima of the scattering loses. 

The width dependence of the transmission parameters for the propagation of the SPP  through 
the layer with higher permittivity,   $B\to A \to B$,
is shown in Fig. \ref{fig51a}. Here, the transmission coefficient exhibits dramatic 
oscillations when the slab thickness varies. 

The  difference between the two cases lies in the role of the plane waves. 
While in the first structure  only a few
plane waves are allowed to propagate through the slab (most of plane waves inside the slab are evanescent), the 
propagation of the SPP through the slab with higher permittivity  
is accompanied by a huge number of plane waves. The transmission  coefficient
of the plane wave $\alpha$ is determined by   the product 
$Lk_{x\alpha}$.  
The radiative loses are enhanced each time when the $x$ component of the wave vector of   one of the plane waves 
fulfills the Fabry-Perot condition for the maximal transmission. 
With many plane waves with different value of $k_x$ inside the slab we observe
huge oscillation of both the transmission coefficient of the SPP and of radiative losses $S$
(not shown in Fig. \ref{fig51a}).
Smooth $L_{\rm slab}$ - dependence of the transmission coefficient is observed only for very thin slabs 
\footnote{
Note that  the wave length $\Lambda_8$ of the SPP is considerably smaller than $2\pi/k_x$ for any accessible plane wave.}.

\begin{figure}[t!]
\begin{center}
\includegraphics[clip,width=11cm]{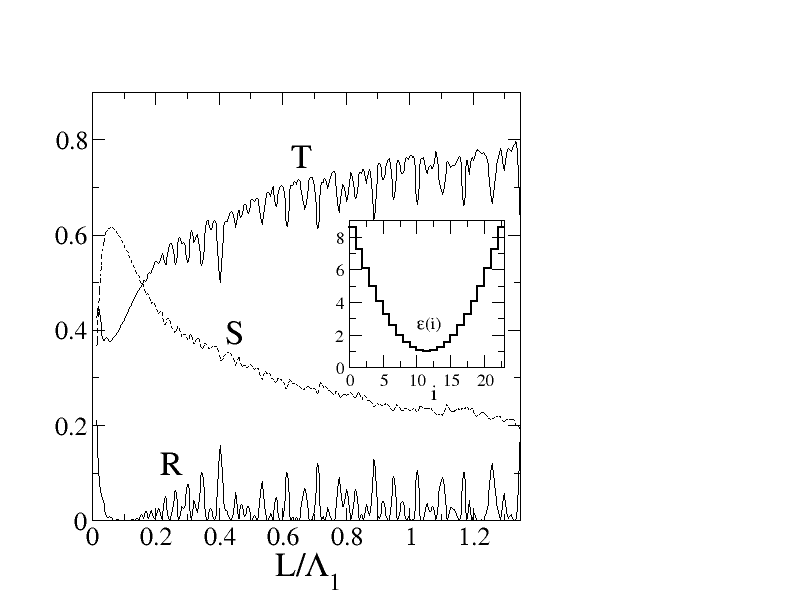} 
\end{center}
\caption{The transmission coefficient $T$,the  reflection coefficient $R$ and radiative  losses $S$
for the normal incidence SPP propagating through the layered structure  with quadratic change  of the dielectric
 permittivity.
Permittivity  $\eps_d = 1 + (2.75 - i/4)^2$ for $i = 0 \dots 22$ is shown in the Inset. The
width of all  layers is same and is shown relative to $\Lambda_1$. 
}
\label{fig3}
\end{figure}
\begin{figure}[t!]
\begin{center}
\includegraphics[clip,width=11cm]{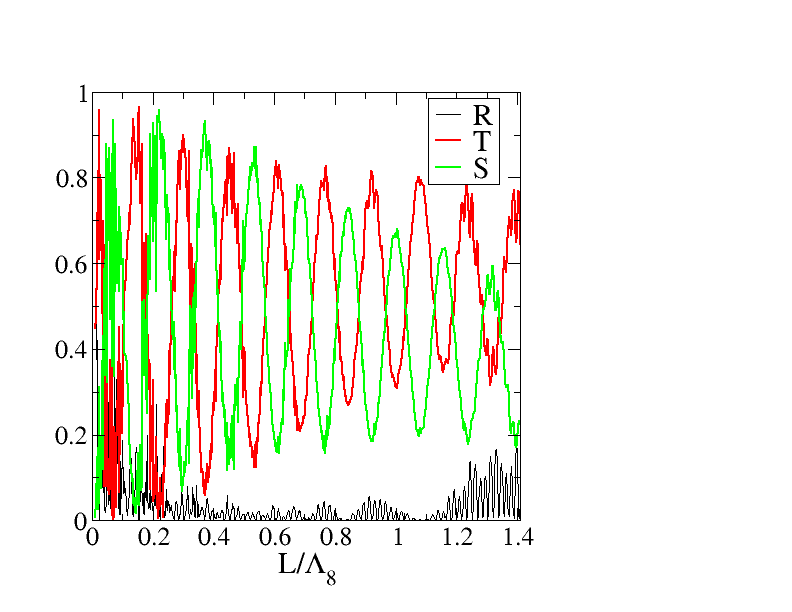}  
\end{center}
\caption{The transmission coefficient $T$, the reflection coefficient $R$ and scattering losses $S$
for the normal incidence SPP propagating through the layered structure with   quadratic change of the 
dielectric permittivity $\eps_d = 8.5625 - (2.75 - i/8)^2$ for $i = 0 \dots 44$. Width of all 
layers is same and is shown relative to $\Lambda_8$. 
}
\label{fig5}
\end{figure}
%
%
%
\section{Quadratic structures}\label{q}
The propagation of the SPP through the slab, discussed in the previous section,
can be enhanced with the use of gradient structures.
As an example, we consider layered structure consisting of $N$ dielectric layers with
permittivities
\be
\eps_i = 1 + (2.75 - i/4)^2,
~~~~~i = 1 \dots N=22.
\ee
The permittivity varies from the maximal value 8.56 to minimal value 1 in the middle of
the structure
(inset of Fig. \ref{fig3}). The width $L$ of each slab  is again considered  relative to the length of the 
 SPP for the dielectric permittivity 
$\eps_d = 1$  ($\Lambda_{1} = 632$ ~nm).
 
The transmission parameters of the SPP are shown in Fig. \ref{fig3}. 
Contrary to the propagation through the slab
 (Fig. \ref{fig51}), 
 no Fabry-Perot oscillations are observable. Instead, we see that
the transmission coefficient  
increases continuously and  the reflection coefficient is almost negligible.

The inverse structure, defined by the quadratic increase of permittivity 
\be
\eps_i = 8.5625 - (2.75 - i/8)^2
~~~~~i = 0 \dots 44,
\ee
exhibits oscillations of the transmission coefficient, 
which are, however, much more moderate than that shown in Fig. \ref{fig51a} for 
a single slab.  The number of oscillations is proportional to  a number of permittivity steps. 
The transmission coefficient  and radiation losses are 
anti-correlated: small radiation losses correspond to 
large transmission coefficient and \textsl{vice versa}.
%
%
%
\begin{figure}[t!]
\begin{center}
\includegraphics[clip,width=11cm]{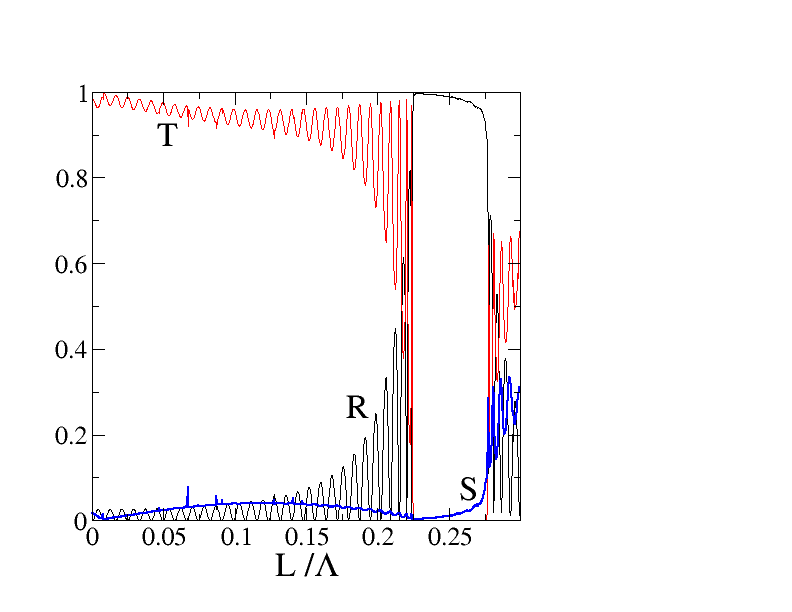} 
\end{center}
\caption{The transmission coefficient  $T$, the  reflection coefficient $R$  and scattering losses $S$
for the normal incidence of the SPP propagating through the
periodic structure of two alternating media as a function of layer 
thickness. Layers are arranged in order $AB AB \dots ABA$, number of periods is $N = 30$.
Permittivities are $\eps_{A} = 1$ and $\eps_{B} = 2$, $N = 30$. Widths  of  layers are
given by Eq. (\ref{QS}). 
}
\label{fig6}
\end{figure}
%
%
%
\begin{figure}[t!]
\begin{center}
\includegraphics[clip,width=11cm]{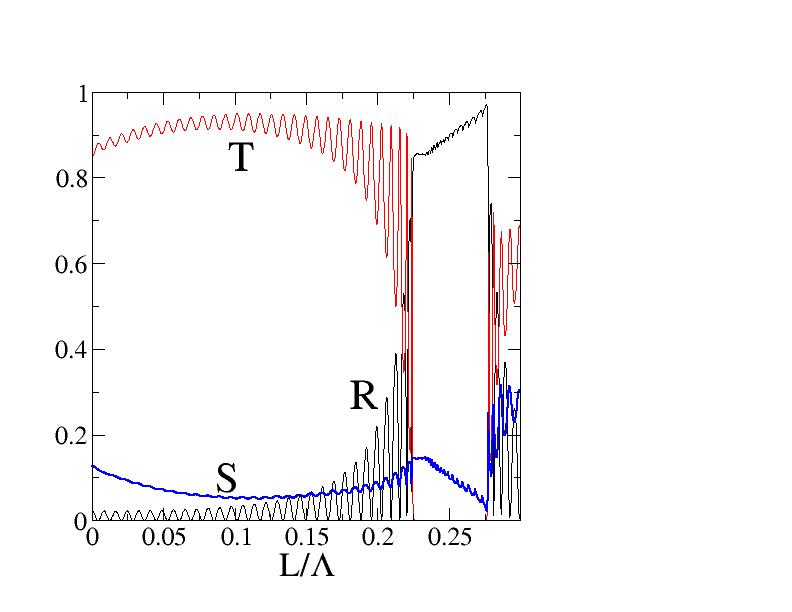}   
\end{center}
\caption{The transmission coefficient $T$, the  reflection coefficient $R$ and scattering losses $S$
for the normal incidence of the SPP propagating through the
periodic structure of two alternating media as a function of layer
thickness. Layers are arranged in order $ABAB \dots AB$, number of periods is $N = 30$.
Permittivities are $\eps_{A} = 1$ and $\eps_{B} = 2$. Widths of  layers are given by Eq. (\ref{QS}).
 Note that the transmission through such structure can be larger than 
the transmission through single interface, $T_0 = 0.86$.
}
\label{fig71}
\end{figure}
\begin{figure}[t!]
\begin{center}
\includegraphics[clip,width=11cm]{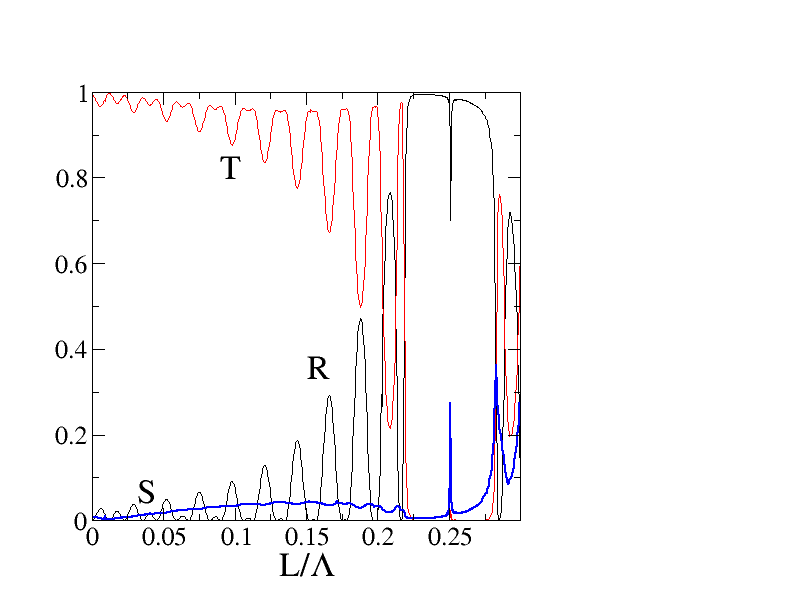} 
\end{center}
\caption{The transmission coefficient $T$, the reflection coefficient $R$ and scattering losses $S$
for the normal incidence of the SPP propagating through the
periodic structure of two alternating media with additional layer $B$ located  in the middle 
of the structure as a function of layer thickness. Layers are arranged in order $ABAB \dots ABBA \dots ABA$, number 
of periods is $N = 22$.
Permittivities are $\eps_{A} = 1$ and $\eps_{B} = 2$. 
}
\label{fig8}
\end{figure}
\begin{figure}[t!]
\begin{center}
\includegraphics[clip,width=11cm]{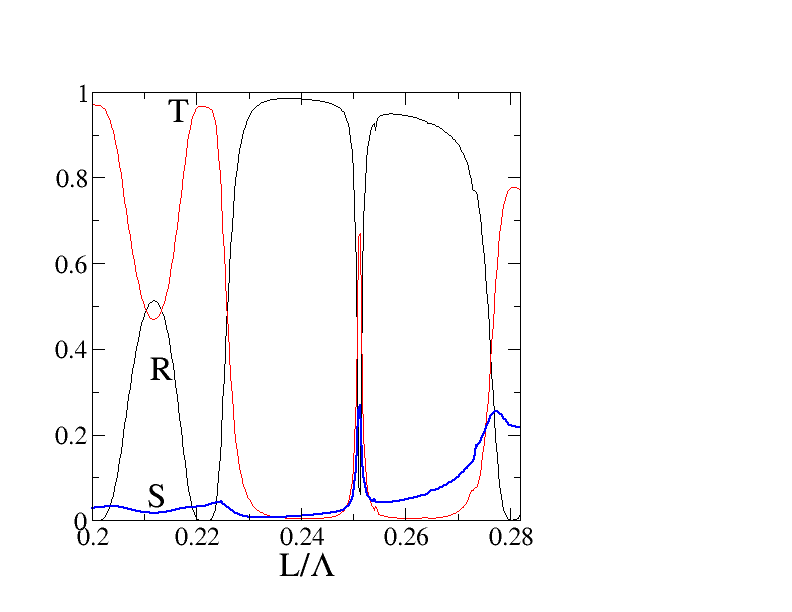} 
\end{center}
\caption{The transmission coefficient $T$, the reflection coefficient $R$ and scattering losses $S$
for the normal incidence of the SPP propagating through the 
periodic structure of two alternating media with additional layer $A$ located  in the middle
of the structure as a function of layer thickness. Layers are arranged in order 
$ABAB \dots BAAB \dots ABA$, number
of periods is $N = 22$.
Permittivities are $\eps_{A} = 3$ and $\eps_{B} = 5$. 
}
\label{fig9}
\end{figure}
%
%

%
\section{Propagation of the SPP through periodic structures}\label{pc}

In this section we consider the propagation of the SPP through 
the layered  structure built by  $N$ periods of layers  
 $ABAB\dots AB$ with permittivities 
$\eps_A$ and $\eps_B$. The periodic structure can be either sandwiched between two semi-infinite media 
$A$ or between medium $A$ and $B$ form the left (right) hand side.

We find that the SPP propagating through such structure can radiate 
less energy than the process of the propagation through a single interface $AB$.
Also,  radiation losses do not increase as the number of stack increases.  
Waves radiated at the first interface are used in the process of scattering at next interface. 

Typical transmission properties of the periodic stack are shown in Figs.   \ref{fig6} and \ref{fig71}.
The structure consists of $N=30$ periods $AB$. 
The width of each layer is  proportional to the wavelength of the SPP:
\be\label{QS}
\ds{\frac{L_A}{L_B}}=\ds{\frac{\Lambda_A}{\Lambda_B}}.
\ee
and permittivities  $\eps_A=1$ and $\eps_B=2$.

Figure \ref{fig6} presents the transmission parameters for the propagation of the SPP through
$N=30$ periods $AB$. Both left and right medium have the permittivity $\eps_A=1$. 
Transmission and reflection coefficients 
exhibit typical oscillations when the layer widths increase.
For $L_A/\Lambda_A>0.22$, we found the gap in which no propagation of the SPP is possible.
Similarly to plane waves, the origin of the transmission gap lies  in the wave character of the SPP.
However, there are two differences between the gap for SPP and for plane waves in one dimensional photonic
crystals.
First, the reflection coefficient of the 
SPP does not reach the unity inside the gap.
A small radiative losses are still observed  since there is
no total gap for the TM polarized plane waves \cite{pcbook}.
Second, since plane waves must assist the propagation of the SPP through each interface inside the 
layered structure,  we expect that the transmission coefficient of the SPP is considerably reduced for layers
widths where the propagation of majority of plane waves is restricted.
Also, the position of the gap for the SPP does not scale with the frequency because of the non-linear frequency dependence of the 
wave length $\Lambda$.

Figure \ref{fig71} shows similar data for the propagation through $N=30$ periods $AB$ but with right medium 
$\eps_B=2$. 
The radiation losses are small for any width of layers.
Plane waves radiated at the first interface into the material $B$  assist the
 transmission of the SPP through next interfaces $BA$
 so  no additional waves should be radiated.
Also, for thin layers,  the transmission coefficient is higher than the transmission coefficient for the single
interface $AB$ ($T_0=0.85$).

The last two Figures, \ref{fig8} and \ref{fig9} demonstrate the transmission parameters
for the SPP propagating through periodic layered
structure with single impurity represented by additional layer $B$ or $A$ in the middle of the structure. Similarly to the 
plane wave photonic crystals, impurity creates isolated level in the band gap, where the
reflection coefficient decreases. Figure \ref{fig8} shows that this decrease of the reflection should not 
automatically imply a good
transmission of the SPP; instead, decrease of the reflection is accompanied with plane wave radiation and the transmission
coefficient is small.
The transmission of SPP is better  for the $A$ impurity layer included into  the layered media with slower permittivity contrast. For instance,
Fig. \ref{fig9} shows well pronounced impurity induced transmission peak ($T\approx 0.7$ inside the gap for 
$\eps_A/\eps_B=3/5$ with impurity $A$ located in the middle of the periodic structure.

\section{Conclusion}

We analyzed the propagation of the surface plasmon polariton - SPP - through various layered structures
in  which the metallic surface is covered by  strips of dielectric materials with different dielectric permittivity.
We  showed that the transmission coefficient of SPP can be considerably increased when the single permittivity step
with high permittivity contrast is substituted by a gradient structure, represented by $N$ thin dielectric slabs with increasing
dielectric permittivity.

We showed that the propagation of the SPP  through more than one dielectric interface must  not necessary require 
an increase of the radiative losses. Plane waves radiated in one interface assist in the process
of the transmission of the SPP through the next interface. 

Although no absorption is considered in our analysis, we expect that 
qualitatively the same results in absorbing structures. Absorption only reduces the transmission coefficient for the SPP
and radiative losses.

\medskip
This work was supported by Project VEGA 0633/09. 

\end{document}